\documentclass[twocolumn,showpacs]{revtex4-1}
\usepackage{graphicx,color}
\usepackage{amsmath,amssymb}
\usepackage{bm}
\usepackage{textcomp}

\begin{document}

\title{Effects of ${\boldsymbol \alpha}$-cluster breaking on 3${\boldsymbol \alpha}$ cluster structures in $^{12}$C}

\author{Tadahiro Suhara}
\affiliation{Matsue College of Technology, Matsue 690-8518, Japan}

\author{Yoshiko Kanada-En'yo}
\affiliation{Department of Physics, Kyoto University, Kyoto 606-8502, Japan}

\date{\today}

\begin{abstract}
To clarify the effects of $\alpha$-cluster breaking on 3$\alpha$ cluster structures in $^{12}$C, we investigate $^{12}$C using a hybrid model that combines the Brink-Bloch cluster model with the $p_{3/2}$ subshell closure wave function. 
We have found that $\alpha$-cluster breaking caused by spin-orbit force significantly changes cluster structures of excited $0^{+}$ states through orthogonality to lower states.
Spatially developed cluster components of the $0^{+}_{2}$ state are reduced.
The $0^{+}_{3}$ state changes from a vibration mode in the bending motion of three $\alpha$ clusters to a chain-like 3$\alpha$ structure having an open triangle configuration.
As a result of these structure changes of $0^{+}$ states, the band assignment for the $2^{+}_{2}$ state is changed by the $\alpha$-cluster breaking effect.
Namely, in model calculations without the $\alpha$-cluster breaking effect, the $0^{+}_{2}$ state is assigned to be the band-head of the $2^{+}_{2}$ state. 
However, when we incorporate $\alpha$-cluster breaking caused by the spin-orbit force, the $0^{+}_{3}$ state is regarded as the band-head of the $2^{+}_{2}$ state.
\end{abstract}


\maketitle

\section{Introduction}\label{introduction}

In nuclear systems, an important feature is the independent single particle motion in a self-consistent mean field.
As a result of spin-orbit splitting coming from the one-body spin-orbit potential, the $jj$-coupling shell model has been able to explain the nature of various nuclei.
Another important feature is the development of clusters.
Usually, well-developed cluster structures appear in excited states close to their cluster decay thresholds, as predicted by the Ikeda's threshold rule~\cite{ikeda_68}.

In light nuclei, it is known that cluster formation occurs in low-lying states and competes with the shell-model structure. 
$^{12}$C is a typical example of this competition~\cite{en'yo_98, itagaki_04, neff_04, en'yo_07, chernykh_07, en'yo_12,fukuoka_13}.
Because the spin-orbit force usually tends to hinder the cluster formation, the ground state of $^{12}$C is not the pure 3$\alpha$ cluster state but is a mixture of 3$\alpha$ cluster structure and $jj$-coupling shell model structure of the $p_{3/2}$ subshell closure; this is supported by the large level spacing between $0^{+}_{1}$ and $2^{+}_{1}$ states.
On the other hand, in excited states of $^{12}$C near the threshold energy, developed 3$\alpha$ cluster structures have been discovered.
It is important to understand the magnitude of this competition and how much it affects the structures of the ground and excited states of $^{12}$C. 
To clarify this problem, it is necessary to have a theoretical model that can describe cluster structures incorporating cluster breaking effects.
Nevertheless, many studies have investigated the cluster structures of $^{12}$C using 3$\alpha$ cluster models without considering cluster breaking.
In the following, we review studies of structures of $^{12}$C on the basis of various theoretical approaches and introduce recent developments in experimental works.

In 3$\alpha$ cluster models, the ground and low-lying states have a compact triangle structure, whereas excited states near and above the 3$\alpha$ cluster threshold energy have well-developed cluster structures~ \cite{uegaki_77, kamimura_77, tohsaki_01, funaki_03, funaki_05, suhara_14, funaki_14}. 
For example, the $0^{+}_2$ state, which is well-known as the Hoyle state, is considered to be an $\alpha$ condensate state in which weakly interacting three $\alpha$ clusters occupy an identical lowest orbit of a mean-field potential~\cite{tohsaki_01, funaki_03, funaki_05}.
This interpretation was indicated by the so-called Tohsaki-Horiuchi-Schuck-R\"{o}pke (THSR) wave function.
The cluster developed structure in the $0^{+}_{2}$ state is also experimentally supported by a quite large monopole transition strength~\cite{yamada_08}.
Another interesting cluster structure was predicted in the $0^{+}_{3}$ state, which is considered to be a vibration mode of acute and obtuse triangle configurations~\cite{uegaki_77}.
However, 3$\alpha$ cluster models cannot describe the detailed properties of ground and excited states of $^{12}$C because they ignore the components of $\alpha$-cluster breaking caused by the spin-orbit force.
For example, the excitation energy of the $2^{+}_{1}$ state is underestimated. 

In the no-core shell model calculation, which is one of the {\it ab initio} calculations, many states of $^{12}$C are reproduced well~\cite{navratil_03, navratil_09}.
However, the model fails to reproduce some states observed near the cluster threshold such as $0^{+}_{2}$ and $0^{+}_{3}$ states.
This seems to be reasonable because these states are cluster developed states in which large proportion of nucleons are distributed over many harmonic oscillator orbits.
To describe these cluster developed states, a model space is required that is significantly larger than those used in the usual no-core shell model calculation~\cite{neff_08}.

Another {\it ab initio} calculation for $^{12}$C has been performed using the chiral effective field theory~\cite{epelbaum_12}.
However, quantitative reproduction of structure properties, such as radii, was not satisfactory even though the calculation reproduced some experimental data such as excitation energies and $E2$ transition strengths. 
For detailed discussion of cluster structures, further developments are required.

The antisymmetrized molecular dynamics (AMD) and the fermionic molecular dynamics (FMD) models have been successful in describing the nature of $^{12}$C from the ground state to the higher excited states~\cite{en'yo_98, itagaki_04, neff_04, en'yo_07, chernykh_07, en'yo_12}.
In these models, the ground and low-lying states are a mixture of cluster and $jj$-coupling shell model configurations.
Because of this mixing, these models reproduce well the excitation energy of the $2^{+}_{1}$ state and $E2$ transition strengths between the low-lying states.
For states near the threshold, these models reproduce the cluster nature, but the detailed cluster structures obtained with these models are somewhat different from those of cluster model results.
For example, the radius of the $0^{+}_{2}$ state can be somewhat smaller than that predicted by the cluster model, although both the models reproduce the monopole transition strength~\cite{suhara_10}.
The $0^{+}_{3}$ state is considered to be a chain-like structure having an obtuse triangle configuration of three $\alpha$ clusters instead of the vibrational state predicted by the cluster model.
The capability of reproducing the experimental data of $^{12}$C is a significant advantage of AMD and FMD.

Recently, several interesting experimental data on $^{12}$C for excited states near the threshold energy have been reported.
A new state, the $2^{+}_{2}$ state at 9.84 MeV, was discovered~\cite{freer_09, itoh_11, zimmerman_13_1}.
The $B(E2)$ transition strength from this $2^{+}_{2}$ state to the ground state was measured to be 1.57$^{+0.14}_{-0.11}$ e$^{2}$fm$^{4}$~\cite{zimmerman_13_2}, which many models fail to reproduce. 
Itoh {\it et al.} found that the broad $0^{+}$ state at 10 MeV consists of two $0^{+}$ states~\cite{itoh_13}.
Theoretically, the existence of two $0^{+}$ states above the $0^{+}_{2}$ state was predicted by Kurokawa and Kat\={o}~\cite{kurokawa_05, kurokawa_07} and also shown by Ohtsubo {\it et al.} \cite{ohtsubo_13} using the orthogonal condition model and the complex scaling method (OCM+CSM).
A band structure including these new states has been an open problem.
In particular, the assignment of the band-head state of the newly measured $2^{+}_{2}$ state is now controversial. 

In this paper, we investigate $^{12}$C with a simple model to clarify $\alpha$-cluster breaking effects on 3$\alpha$ cluster structures.
In this model, we superpose the Brink-Bloch (BB) cluster model wave functions and the $p_{3/2}$ subshell closure wave function to incorporate the mixing of cluster breaking components in cluster structures. 
This model is regarded as an extension of the generator coordinate method of a cluster model. 
The difference from traditional cluster models is the mixing of the $p_{3/2}$ subshell closure wave function, which is the lowest configuration of the $jj$-coupling shell model.
The results obtained with this model are consistent with the AMD and FMD results even though it is much simpler than those models.
In this model space, we changed the strength of the spin-orbit force by hand to control the mixing of the $\alpha$-cluster breaking component, i.e., the $p_{3/2}$ subshell closure wave function.
We found that the mixing of the $\alpha$-cluster breaking component significantly changes the cluster configurations and band structure of excited states.

This paper is organized as follows. 
In Sec.~\ref{formulation}, we explain the wave function and Hamiltonian used in the present study. 
In Sec.~\ref{results}, we show calculated results.
In Sec.~\ref{discussion}, we discuss differences in cluster configurations in $0^{+}$ states and band structure between calculations with and without $\alpha$-cluster breaking.
We compare our results with other theoretical calculations.
Finally, we summarize the findings of the present study in Sec.~\ref{summary}.

\section{Formulation}\label{formulation}

\subsection{Wave function}

\begin{figure}[tb]
	\centering
	\includegraphics[width = 5.5 cm]{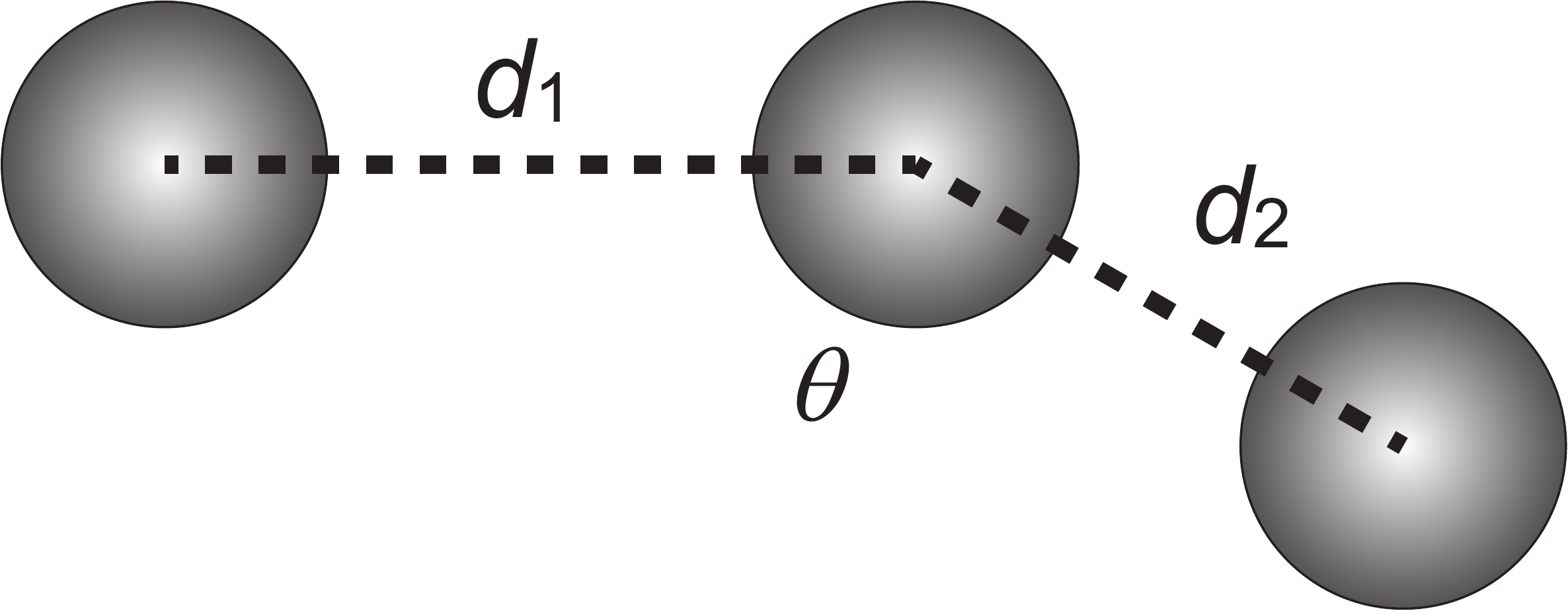}
	\caption{Schematic of a 3$\alpha$ cluster structure.
	The configuration is characterized by the distances between $\alpha$ clusters, $d_{1}$ and $d_{2}$, and the bending angle, $\theta$.}
	\label{fig:configuration}
\end{figure}

In this work, we investigate $^{12}$C using a simple model that combines the BB cluster model with the lowest wave function of the $jj$-coupling shell model.
We superpose many 3$\alpha$ BB wave functions and the single $p_{3/2}$ subshell closure wave function.

The BB wave function~\cite{brink_66} for three $\alpha$ clusters is written as   \begin{equation}
	\Phi^{({\rm BB})}(\bm{R}_{1}, \bm{R}_{2}, \bm{R}_{3})= {\cal A}[\varphi^\alpha_{1}, \varphi^\alpha_{2}, \varphi^\alpha_{3}],
\end{equation}
\begin{equation}
	\varphi^\alpha_{i} = \phi_{\bm{R}_{i}} \phi_{\bm{R}_{i}} \phi_{\bm{R}_{i}} \phi_{\bm{R}_{i}} \chi_{p \uparrow} \chi_{p \downarrow} \chi_{n \uparrow} \chi_{n \downarrow}, 
\end{equation}
\begin{equation}
	\phi_{\bm{R}_{i}} = \left( \frac{2 \nu}{\pi} \right)^{3/4} \exp \left[ - \nu (\bm{r} - \bm{R}_{i})^{2} \right], 
\label{eq:brink_single}
\end{equation}
where $\bm{R}_{i}$ is a real vector that indicates the position of the $i$th $\alpha$ cluster.
For the width parameter, we use the value $\nu = 0.235$ fm$^{-2}$ same as that used in the previous study of $^{12}$C using AMD~\cite{suhara_10}.
This parameter was originally from a variational calculation for the ground state of $^{9}$Be in Ref.~\cite{okabe_79}.
The configuration of the BB wave function is characterized by the distances between $\alpha$ clusters, $d_{1}$ and $d_{2}$, and the bending angle, $\theta$, as shown in Fig.~\ref{fig:configuration}.
Because of symmetry, we can limit distances to $d_{1} \le d_{2} $ without loss of generality.
To describe relative motions of clusters sufficiently, we generate 252 BB wave functions by taking the distance $d_{i} = 1, 2, \cdots, 6$ fm and the bending angle $\theta = n \pi /12$ with $n = 1, 2, \cdots, 12$.

To describe the $p_{3/2}$ subshell closure wave function, we adopted the method proposed in Refs.~\cite{itagaki_05,suhara_13}. 
In this method, the single particle wave function is described by a Gaussian wave packet in the same manner as in the BB wave function Eq.~\eqref{eq:brink_single}.
However, the center of the wave packet for the $j$th nucleon belonging to the $i$th $\alpha$ cluster is replaced by a complex parameter $\bm{\zeta}_{j}$,
\begin{equation}
	\bm{\zeta}_{j} = \bm{R}_{i} + i \bm{e}^{\text{spin}}_{j} \times \bm{R}_{i},
\end{equation}
where $\bm{e}^{\text{spin}}_{j}$ is a unit vector for the intrinsic spin direction of the $j$th nucleon.
In the limit of $\bm{R}_{i} \rightarrow 0$, this wave function describes the $p_{3/2}$ subshell closure wave function.
For the width parameter of the $p_{3/2}$ subshell closure wave function, we use the same value $\nu = 0.235$ fm$^{-2}$ of the BB wave function to exactly separate the center-of-mass motion from the total wave function.

As the final wave function, we superpose the 252 3$\alpha$ BB wave functions and the single $p_{3/2}$ subshell closure wave function with the parity and angular momentum projection,
\begin{equation}
	| \Psi_{J^{\pi}} \rangle = \sum_{i, K} c_{iK}^{J^{\pi}} \hat{P}^{J^{\pi}}_{MK} | \Phi^{({\rm BB})}_{i} \rangle + c_{p_{3/2}} \hat{P}^{J^{\pi}}_{MK} | p_{3/2} \rangle,
	\label{eq:final}
\end{equation}
where $\hat{P}^{J^{\pi}}_{MK}$ is the parity and angular momentum projection operator.
The coefficients are determined by the diagonalization of the norm and Hamiltonian matrices.

The superposition of BB wave functions without the $p_{3/2}$ subshell closure wave function corresponds to the traditional 3$\alpha$ cluster GCM calculation without $\alpha$-cluster breaking. 
In the present model, $\alpha$-cluster breaking caused by the spin-orbit force is incorporated by adding the $p_{3/2}$ subshell closure wave function to the 3$\alpha$ cluster model space. 
Note that the final wave functions for the non-zero angular momentum states are the same as those obtained by the 3$\alpha$ cluster GCM calculation because the single $p_{3/2}$ subshell closure wave function is a $J^{\pi} = 0^{+}$ eigenstate; therefore, $\hat{P}^{J^{\pi}}_{MK} | p_{3/2} \rangle$ vanishes for $J^{\pi} \ne 0^{+}$. 
In other words, the $\alpha$-cluster breaking caused by the spin-orbit force affects only the $0^{+}$ states through the mixing of the $p_{3/2}$ subshell closure wave function.

In the present work, we omit cluster breaking components for the $J^{\pi} \ne 0^{+}$ states because they give only minor effect compared with the significant cluster breaking effects in the $J^{\pi} = 0^{+}$ states.
Indeed, the structure of the cluster developed states does not change qualitatively even if we adopt the cluster breaking components for the $J^{\pi} \ne 0^{+}$ states.
To simplify the discussion, we adopt only the $p_{3/2}$ subshell closure wave function, which gives the major contribution of the $\alpha$-cluster breaking effect.

\subsection{Hamiltonian}

The Hamiltonian consists of the kinetic energy $\hat{t}_i$, effective central force $\hat{V}^{\text{cent.}}_{ij}$, effective spin-orbit force $\hat{V}^{\text{LS}}_{ij}$, and Coulomb force $\hat{V}^{\text{Coul.}}_{ij}$,
\begin{equation}
	\hat{H}= \sum_{i} \hat{t}_i - \hat{T}_{G} + \sum_{i<j} \hat{V}^{\text{cent.}}_{ij}
		 + \sum_{i<j} \hat{V}^{\text{LS}}_{ij} + \sum_{i<j} \hat{V}_{ij}^{\text{Coul.}}, 
\label{eq:hml}
\end{equation}
where the center-of-mass kinetic energy $\hat{T}_{G}$ is subtracted. 
For the effective central force, we used the Volkov No.~2 force~\cite{volkov_65} with Majorana parameter $M=0.60$. 
For the effective spin-orbit force, we used the spin-orbit term of the G3RS interaction~\cite{yamaguchi_79},
\begin{equation}
	\hat{V}^{\text{LS}}_{ij} = \sum_{k=1}^{2} u_{k} \exp \left[- \left( \frac{\hat{r}_{ij}}{b_{k}} \right)^{2} \right] \hat{P}(^{3} \text{O}) \hat{\bm{L}} \cdot \hat{\bm{S}},
\end{equation}
where $\hat{P}(^{3} \text{O})$ is a projection operator onto the triplet odd state.
We chose $u_{1} = - u_{2} \equiv u_{\rm ls}$, where $u_{\rm ls}$ is the strength of the spin-orbit force.
We used the same interaction with $u_{\rm ls} = 1600$ MeV as that used in a previous study of $^{12}$C using AMD~\cite{suhara_10}.
We also investigate the $u_{\rm ls}$ dependence of energy levels and the degree of $\alpha$-cluster breaking by changing the strength $u_{\rm ls}$ from 0 MeV to 3200 MeV.

\section{Results}\label{results}

\begin{figure}[tb]
\centering
\includegraphics[width = 8.6cm]{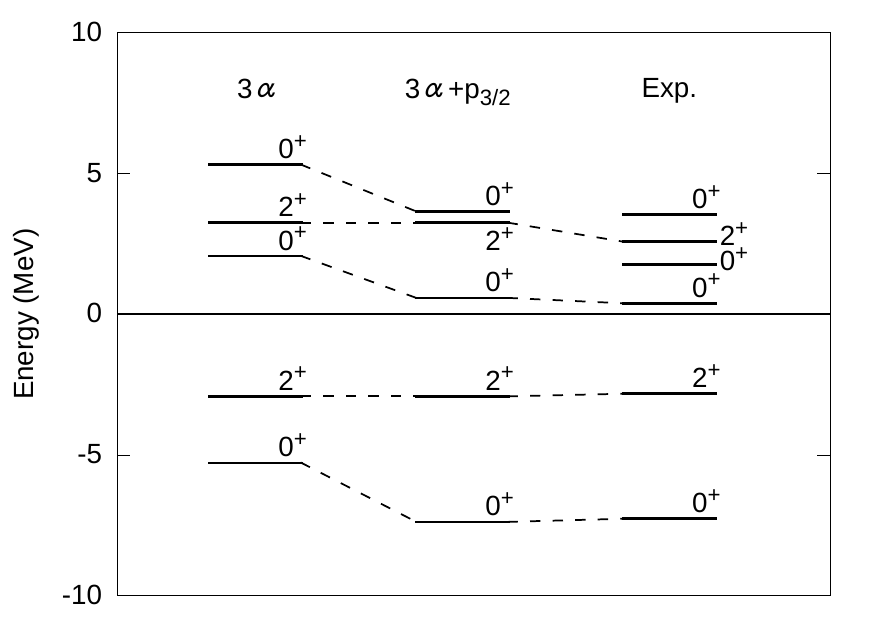}
\caption{Comparison of calculated energy levels of $0^{+}$ and $2^{+}$ states with experimental ones.
Energies are measured from the $3\alpha$ threshold energy.
Theoretical and experimental threshold energies are $-82.9$ MeV and $-84.9$ MeV, respectively. 
Levels labeled as ``3$\alpha$'' and ``3$\alpha$+$p_{3/2}$'' show cluster model results and present results for $u_{\rm ls} = 1600$ MeV, respectively.}
\label{fig:levels_0+_2+}
\end{figure}

In Fig.~\ref{fig:levels_0+_2+}, we show the energy levels of $0^{+}$ and $2^{+}$ states calculated with the present model using $u_{\rm ls} = 1600$ MeV, which is the same strength used in our previous study of $^{12}$C using AMD~\cite{suhara_10}, and those obtained with the $3\alpha$-cluster GCM calculation. 
The former and the latter correspond to calculations with and without $\alpha$-cluster breaking, i.e., the mixing of the $p_{3/2}$ subshell closure wave function caused by the spin-orbit force, which we call ``3$\alpha$+$p_{3/2}$'' and ``3$\alpha$'', respectively. 
Experimental data are taken from Refs.~\cite{itoh_11, itoh_13, selove_90}.
Energies are measured from the $3\alpha$ threshold energy.
The theoretical and experimental threshold energies are $-82.9$ MeV and $-84.9$ MeV, respectively.

In the ``3$\alpha$'' case, the spin-orbit force provides zero energy contribution because the expectation value of the spin-orbit force vanishes in the pure $3\alpha$ cluster model space. 
In the ``3$\alpha$+$p_{3/2}$'' result, the $2^{+}$ states are completely the same as those of the ``3$\alpha$'' result, but the $0^{+}$ energies are decreased by the spin-orbit force because of the mixing of the $p_{3/2}$ subshell closure wave function.
The results for ``3$\alpha$+$p_{3/2}$'' agree with experimental results very well except for the absence of a $0^{+}$ state observed around 10 MeV.
In particular, the level spacing between $0^{+}_{1}$ and $2^{+}_{1}$ states of ``3$\alpha$+$p_{3/2}$'' agrees well with the experimental one, which is largely underestimated in ``3$\alpha$''. 
This large spacing comes from the energy gain of the spin-orbit force in the $0^{+}_{1}$ state with the mixing of the $p_{3/2}$ subshell closure wave function.
Any microscopic $\alpha$ cluster model fails to reproduce this large level spacing, and therefore, this is an evidence for $\alpha$-cluster breaking.

As shown later, the ``3$\alpha$+$p_{3/2}$'' result is qualitatively consistent with the AMD and FMD results for $^{12}$C. 
This implies that the major difference between the AMD and FMD results and 3$\alpha$ cluster model calculations is the $\alpha$-cluster breaking effect, which can be qualitatively simulated by the mixing of only one configuration of the $p_{3/2}$ subshell closure to the $3\alpha$ cluster model space in the present model.
In Sec.~\ref{discussion}, we discuss in detail the structure differences between $0^{+}$ states with and without  $\alpha$-cluster breaking.

\begin{figure}[tb]
	\centering
	\includegraphics[width = 8.6cm]{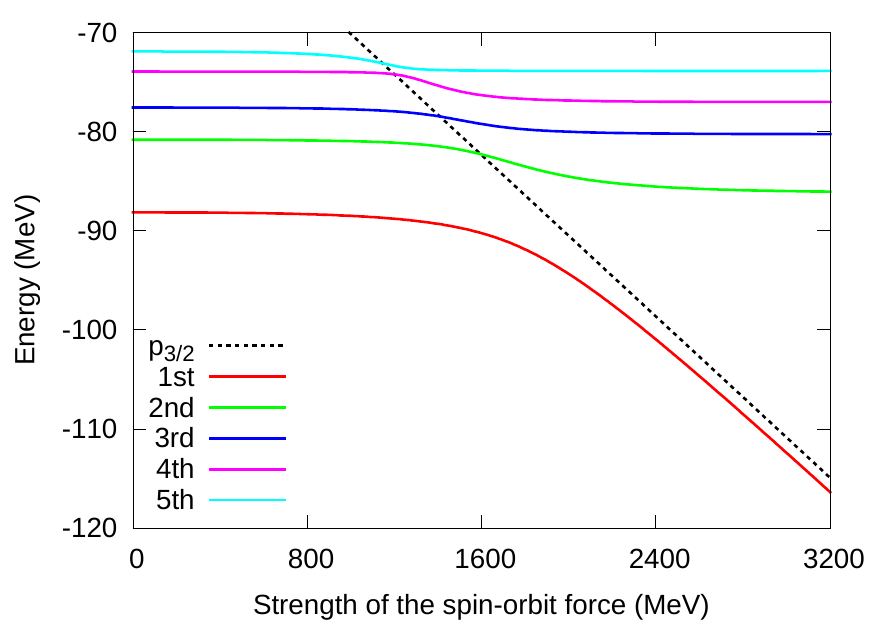}
	\caption{(color online).
	Change in energy levels of $0^{+}$ states against the strength of the spin-orbit force $u_{\rm ls}$.
	Dotted line is the energy of the $p_{3/2}$ subshell closure wave function.}
\label{fig:level_LS_dependence}
\end{figure}

As mentioned above, $\alpha$-cluster breaking caused by the spin-orbit force gives a significant effect in $^{12}$C. 
To clarify the mechanism of the mixing of the $\alpha$-cluster breaking component, we consider the dependence of $0^{+}$ states on the strength of the spin-orbit force by changing the strength parameter $u_{\rm ls}$ from 0 MeV to 3200 MeV in the present model.
The calculation using $u_{\rm ls} = 0$ MeV is equivalent to the ``3$\alpha$'' calculation.
As the strength of the spin-orbit force increases, $0^{+}$ energy levels change due to $\alpha$-cluster breaking, i.e., the $p_{3/2}$ subshell closure wave function.

In Fig.~\ref{fig:level_LS_dependence}, we show the change in energy levels of $0^{+}$ states against the strength of the spin-orbit force $u_{\rm ls}$. 
The energy of the $p_{3/2}$ subshell closure wave function is also shown for reference.
For $u_{\rm ls} \le 800$ MeV, the energies are almost independent of the strength of the spin-orbit force.
In this region, the energy levels and wave functions themselves are almost equal to those obtained from the pure cluster model calculation without $\alpha$-cluster breaking.
When $u_{\rm ls} \ge 2400$ MeV, the $0^{+}$ levels except for the lowest one do not depend on the strength of the spin-orbit force; however, the energy of the lowest $0^{+}$ state linearly decreases and shows almost the same $u_{\rm ls}$ dependence as that of the $p_{3/2}$ subshell closure state. 
These imply that structures of the states do not change when the strength of the spin-orbit force is extremely weak or strong.
However, around $u_{\rm ls} = 1600$ MeV, energy levels depend on the strength $u_{\rm ls}$ because of level crossings of the cluster states and the $p_{3/2}$ subshell closure wave function. 
In this transient region, a change in the wave function occurs for each state.

\begin{table}[tb]
	\caption{Percentages of squared overlaps between $0^{+}$ GCM wave functions and the $p_{3/2}$ subshell closure wave function for $u_{\rm ls} = 0, 800, 1600, 2400, 3200$ MeV.}
	\label{table:squared_overlap_p3/2}
	\centering
	\begin{tabular}[t]{ccccccc} \hline \hline
	& \multicolumn{5}{c}{$u_{\rm ls}$ (MeV)} & \\ \cline{2-6} 
	State  & 0 & 800 & 1600 & 2400 & 3200 \\ \hline
	$0^{+}_{1}$ & 2.46 & 6.11 & 34.3 & 89.6 & 97.0 \\
	$0^{+}_{2}$ & 0.51 & 1.86 & 25.9 & 6.08 & 1.14 \\
	$0^{+}_{3}$ & 0.43 & 1.96 & 16.8 & 1.13 & 0.25 \\
	$0^{+}_{4}$ & 0.07 & 0.54 & 14.0 & 0.93 & 0.25 \\
	$0^{+}_{5}$ & 0.68 & 5.82 & 0.71 & 0.25 & 0.02 \\
	\hline \hline
	\end{tabular}
\end{table}

In Table~\ref{table:squared_overlap_p3/2}, we show the squared overlaps between the $0^{+}$ GCM wave functions and the $p_{3/2}$ subshell closure wave function,
\begin{equation}
	O_{p_{3/2}} = | \langle \Psi_{0^{+}} | p_{3/2} \rangle |^{2},
\end{equation}
for $u_{\rm ls} = 0, 800, 1600, 2400, 3200$ MeV.
These overlaps measure the magnitude of $\alpha$-cluster breaking.
For $u_{\rm ls} \le 800$ MeV, there is little, if any, $\alpha$-cluster breaking in $0^{+}$ states. 
In particular, for $u_{\rm ls} = 0$ MeV, $0^{+}$ states are composed of pure 3$\alpha$-cluster components decoupled from the $\alpha$-clusters breaking component. 
Note that the $p_{3/2}$ subshell closure wave function is not orthogonal to the 3$\alpha$ cluster wave functions but slightly overlaps with the $0^{+}$ projected $SU(3)$-limit 3$\alpha$ cluster wave function,
\begin{equation}
	| \langle p_{3/2} | \hat{P}^{0+}_{00} | SU(3) \rangle |^{2} = 5/81.
\end{equation}
Therefore, even for $u_{\rm ls} = 0$ MeV without $\alpha$-cluster breaking, the squared overlap of the $0^{+}_{1}$ state is not zero but finite of the same order as 5/81. 
For $u_{\rm ls} \ge 2400$ MeV, the $0^{+}_{1}$ state is dominated by the $p_{3/2}$ subshell closure wave function.
This implies that $\alpha$-clusters are broken almost completely in the lowest state.
For higher states, there is no $\alpha$-cluster breaking.
For $u_{\rm ls} = 1600$ MeV, the squared overlaps for $0^{+}_{1,2,3,4}$ states are 10 - 30 \%, which implies that significant components of $\alpha$-clusters are broken in these states.
In other words, 3$\alpha$ cluster states couple significantly with the cluster breaking configuration in these states. 

\begin{table}[tb]
	\caption{Rms radii of $0^{+}$ states.
	Experimental data are from Refs.~\cite{ozawa_01, danilov_09}.
	The unit is fm.}
	\label{table:radii}
	\centering
	\begin{tabular}[t]{ccccccc} \hline \hline
	& \multicolumn{5}{c}{$u_{\rm ls}$ (MeV)} & \\ \cline{2-6} 
	State  & 0 & 800 & 1600 & 2400 & 3200 & Exp. \\ \hline
	$0^{+}_{1}$ & 2.53 & 2.51 & 2.35 & 2.11 & 2.09 & $2.35 \pm 0.02$ \\
	$0^{+}_{2}$ & 3.44 & 3.43 & 2.99 & 2.70 & 2.69 & $2.89 \pm 0.04$ \\
	$0^{+}_{3}$ & 3.63 & 3.62 & 3.39 & 3.50 & 3.51 & \\
	$0^{+}_{4}$ & 3.76 & 3.74 & 3.40 & 3.60 & 3.62 & \\
	$0^{+}_{5}$ & 3.76 & 3.74 & 3.83 & 3.81 & 3.81 & \\
	\hline \hline
	\end{tabular}
\end{table}

In Table~\ref{table:radii}, we show the root-mean-square (rms) radii of the $0^{+}$ states for $u_{\rm ls} = 0, 800, 1600, 2400, 3200$ MeV.
Reproduction of experimental values is fairly good at $u_{\rm ls} = 1600$ MeV.
As the strength of the spin-orbit force increases, the radii of the $0^{+}_{1}$ and $0^{+}_{2}$ states decrease.
The shrinking mechanism differs between $0^{+}_{1}$ and $0^{+}_{2}$ states. 
For the $0^{+}_{1}$ state, the shrinking comes from the decreasing of 3$\alpha$ cluster components by the significant mixing of the $p_{3/2}$ subshell closure wave function.
However, for the $0^{+}_{2}$ state, this decrease mainly comes from the mixing of relatively compact cluster components because of orthogonality to the $0^{+}_{1}$ state.
As the strength of the spin-orbit force increases, the 3$\alpha$ component decreases in the $0^{+}_{1}$ state, which makes the $0^{+}_{2}$ state includes more compact 3$\alpha$ cluster components.
Details are discussed in Sec.~\ref{discussion}.

\begin{table}[tb]
	\caption{Monopole transition strengths between $0^{+}$ states. 
	Experimental data are from Ref.~\cite{selove_90}.
	The unit is fm$^{2}$.}
	\label{table:monopole}
	\centering
	\begin{tabular}[t]{ccccccc} \hline \hline
	& \multicolumn{5}{c}{$u_{\rm ls}$ (MeV)} & \\ \cline{2-6} 
	Transition  & 0 & 800 & 1600 & 2400 & 3200 & Exp. \\ \hline
	$0^{+}_{1} \rightarrow 0^{+}_{2}$ & 8.2 & 8.1 & 8.1 & 2.8 & 1.3 & $5.4 \pm 0.2$ \\
	$0^{+}_{1} \rightarrow 0^{+}_{3}$ & 5.5 & 5.4 & 2.1 & 0.5 & 0.4 & \\
	$0^{+}_{2} \rightarrow 0^{+}_{3}$ & 10.7 & 10.9 & 16.8 & 10.6 & 9.7 & \\
	\hline \hline
	\end{tabular}
\end{table}

In Table~\ref{table:monopole}, we show the monopole transition strengths between $0^{+}$ states.
The transition strengths rapidly change around $u_{\rm ls} = 1600$ MeV.
The $0^{+}_{1} \rightarrow 0^{+}_{2}$ transition strength for the strong spin-orbit force is smaller than that for the weak spin-orbit force.
This result is consistent with the fact that the $0^{+}_{1}$ state becomes dominated by the $p_{3/2}$ subshell closure wave function and loses the cluster correlation, which enhances the monopole transition strength to the cluster developed states. 

\begin{table}[tb]
	\caption{$E2$ transition strengths from $2^{+}$ states to $0^{+}$ states.
	Experimental data are from Refs.~\cite{selove_90, zimmerman_13_2}.
	The unit is $e^{2}$fm$^{4}$.}
	\label{table:E2}
	\centering
	\begin{tabular}[t]{ccccccc} \hline \hline
	& \multicolumn{5}{c}{$u_{\rm ls}$ (MeV)} & \\ \cline{2-6}
	Transition & 0 & 800 & 1600 & 2400 & 3200 & Exp. \\ \hline
	$2^{+}_{1} \rightarrow 0^{+}_{1}$ & 10.8 & 10.6 & 7.4& 1.2 & 0.4 & $7.6 \pm 0.4$ \\
	$2^{+}_{1} \rightarrow 0^{+}_{2}$ & 1.4 & 1.6 & 5.1 & 11.1 & 11.8 & $2.6 \pm 0.4$ \\
	$2^{+}_{1} \rightarrow 0^{+}_{3}$ & 0.4 & 0.4 & 0.2 & 0.2 & 0.3 & \\
	$2^{+}_{2} \rightarrow 0^{+}_{1}$ & 4.0 & 3.6 & 1.1 & 0.0 & 0.0 & 1.57$^{+0.14}_{-0.11}$ \\
	$2^{+}_{2} \rightarrow 0^{+}_{2}$ & 183 & 179 & 76.5 & 14.8 & 12.0 & \\
	$2^{+}_{2} \rightarrow 0^{+}_{3}$ & 64.4 & 69.8 & 166 & 206 & 207 & \\
	\hline \hline
	\end{tabular}
\end{table}

The mixing of the $\alpha$-cluster breaking in $0^{+}$ states also affects the $E2$ transition strengths between $0^{+}$ and $2^{+}$ states.
The calculated $E2$ transition strengths from $2^{+}$ states to $0^{+}$ states are shown in Table~\ref{table:E2}.
Similar to the monopole transition strengths, the $E2$ transition strengths change drastically around $u_{\rm ls} = 1600$ MeV.
The agreement between theoretical and experimental results is rather good around $u_{\rm ls} = 1600$ MeV, but it is worse in both the strong and weak limits of the strength of spin-orbit force.
Considering the overall results presented up to here, reproduction of experimental values is better around $u_{\rm ls} = 1600$ MeV, where the degree of mixing of cluster and cluster breaking components is significant not only in the ground state but also in excited $0^{+}$ states.

The $E2$ transition strengths from the $2^{+}_{2}$ state are important for determining the band assignment for this state, which has been attracting significant interest. 
Its $E2$ transition strengths to the $0^{+}_{2, 3}$ states are remarkably large.
Interestingly, the behavior of the magnitude is reversed by $\alpha$-cluster breaking in the change from $u_{\rm ls} = 800$ MeV to $1600$ MeV.
For $u_{\rm ls} = 0$ MeV in the weak limit of the spin-orbit force, the $2^{+}_{2} \rightarrow 0^{+}_{2}$ transition strength is almost three times larger than the $2^{+}_{2} \rightarrow 0^{+}_{3}$ one.
However, for $u_{\rm ls} = 1600$ MeV, the $2^{+}_{2} \rightarrow 0^{+}_{3}$ transition strength is larger by a factor of two than the $2^{+}_{2} \rightarrow 0^{+}_{2}$ one.
The strongest $E2$ transition to the $0^{+}_{3}$ state for $u_{\rm ls} = 1600$ MeV, in which the reproduction of experimental results is good, indicates that the $0^{+}_{3}$ state is likely to be the band-head of the $2^{+}_{2}$ instead of the $0^{+}_{2}$ state.

The present results indicate that the band structure of higher excited states is affected by $\alpha$-cluster breaking caused by the spin-orbit force. 
Even though the $0^{+}_{2, 3}$ states for $u_{\rm ls} = 1600$ MeV still contain dominant $3\alpha$ cluster configurations, their $E2$ transition strengths from the $2^{+}_{2}$ state are considerably different from those with $u_{\rm ls} = 0$ and 800 MeV for the calculation without $\alpha$-cluster breaking. 
The critical effect of $\alpha$-cluster breaking on the $2^{+}_{2}$ band assignment mainly originates from the change of 3$\alpha$ configurations in the cluster model space.
More details of the cluster structure change of the $0^{+}_{2}$ and $0^{+}_{3}$ states caused by $\alpha$-cluster breaking are discussed in the next section.

\section{Discussion}\label{discussion}

\subsection{Structure changes of $0^{+}$ states}

\begin{figure}[tb]
	\centering
	\includegraphics[width = 8.6cm]{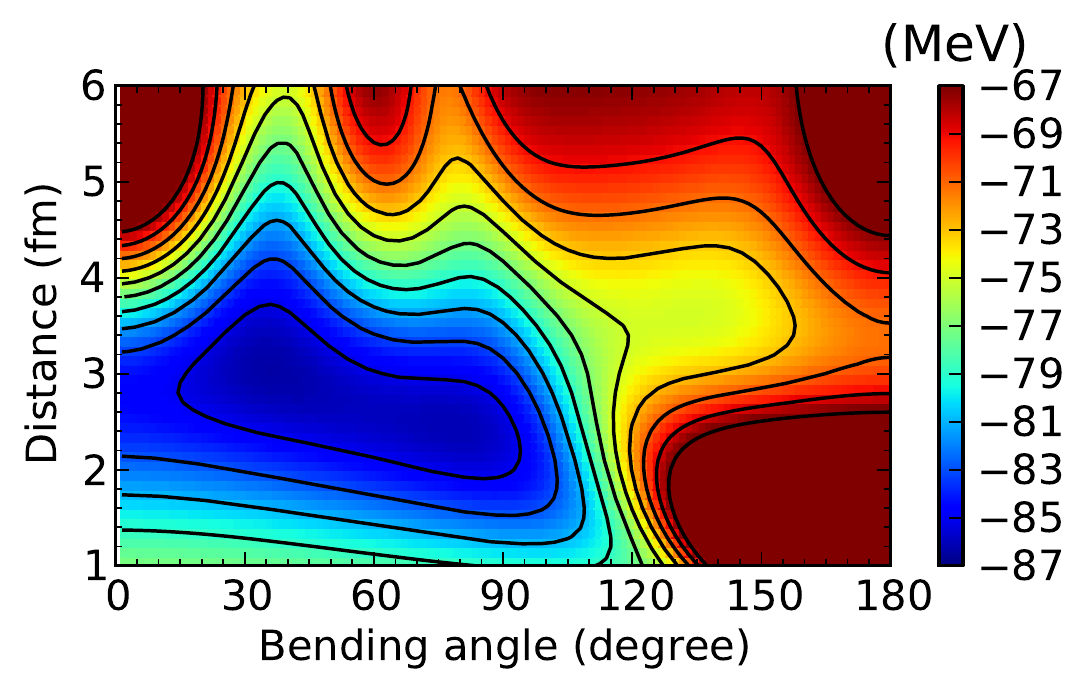}
	\caption{(color online).
	$0^{+}$ energy surface of a single BB wave function for 3$\alpha$ clusters in which equal distances between $\alpha$ clusters are assumed, i.e., $d_{1} = d_{2} \equiv d$.}
	\label{fig:energy_surface}
\end{figure}

To investigate the structure changes of $0^{+}$ states by $\alpha$-cluster breaking, we focus on 3$\alpha$ configurations contained in $0^{+}$ states obtained from calculations with and without $\alpha$-cluster breaking.
We show the energy surface for 3$\alpha$ configurations in Fig.~\ref{fig:energy_surface}. 
The figure shows the $0^{+}$ energy of a single BB wave function for the isosceles triangle configuration (equal distances, $d_{1} = d_{2} \equiv d$) on the two-dimensional plane for distance and angle.
Note that the spin-orbit force has no energy contribution in the 3$\alpha$ subspace.

Around $(d, \theta) = (2.5 \text{ fm}, 60^{\circ})$, a rather deep minimum exists, which corresponds to the ground state of $^{12}$C in the pure 3$\alpha$ model space.
The 3$\alpha$ cluster structure in this region with finite distances has an equilateral triangle shape and has a spatial development to some extent.
This indicates that higher cluster correlation beyond the SU(3) shell-model limit is contained even in the ground state.
Around $(d, \theta) = (4.0 \text{ fm}, 130^{\circ})$ for obtuse (open) triangle structures, a plateau with an energy of $-74$ MeV exists and continues into the area of small bending angles for acute triangle structures.
In the 3$\alpha$ cluster model calculation without $\alpha$-cluster breaking, $\alpha$ clusters move almost freely on this soft mode of energy surface involving various triangle configuration, and construct the $0^{+}_{2}$ state, which is a typical example of ``gaslike'' cluster states.
The $0^{+}_{3}$ state as the vibrational excitation is built on top of the $0^{+}_{2}$ state along this soft energy surface, as suggested by Uegaki {\it et al.}~\cite{uegaki_77}.
This situation changes somewhat when $\alpha$-cluster breaking occurs.

\begin{figure*}[tb]
	\centering
	\includegraphics[width = \linewidth]{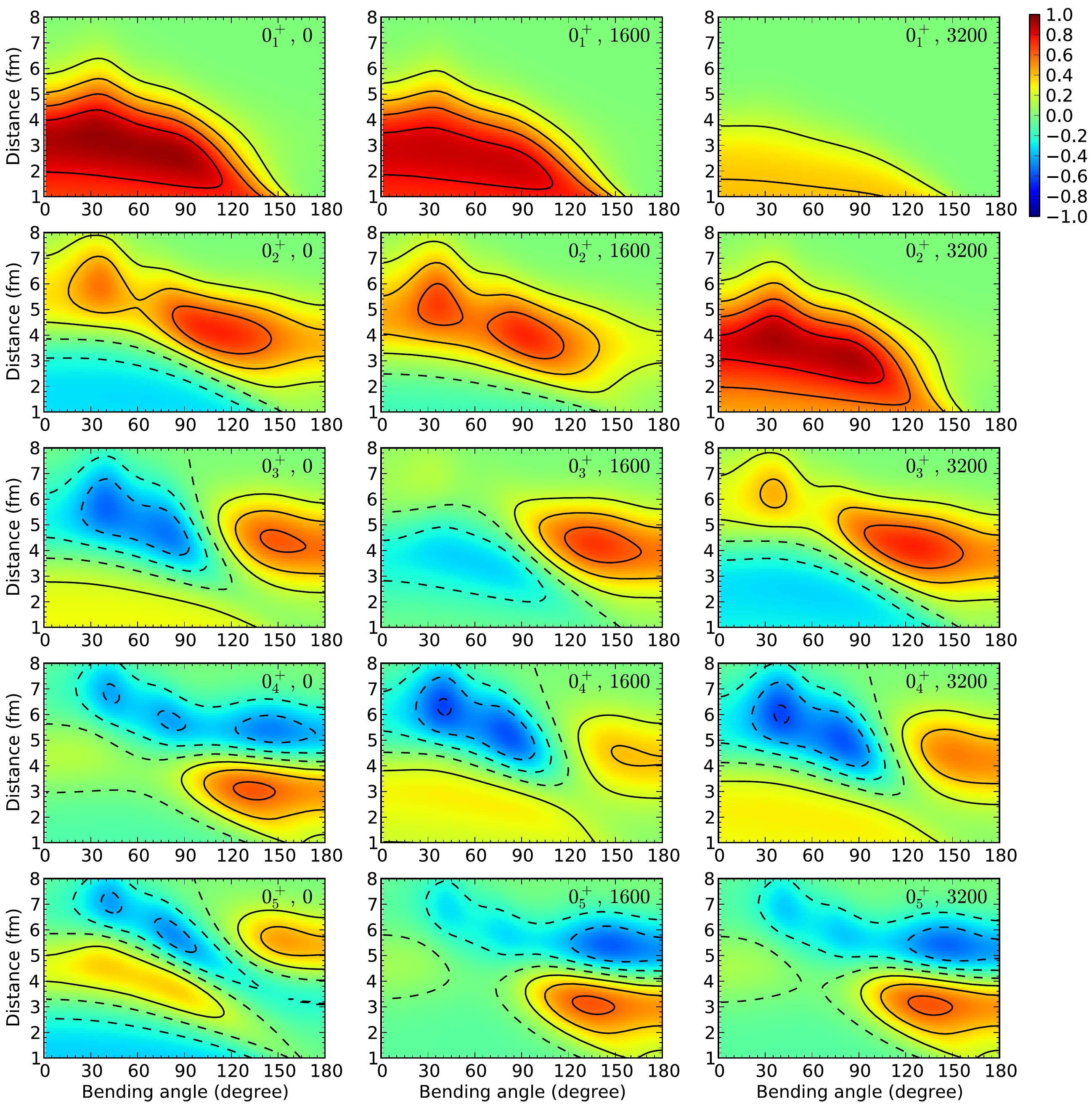}	
	\caption{(color online).
	Overlap surfaces between $0^{+}$ GCM states and single BB wave functions for $u_{\rm ls} = 0, 1600, 3200$ MeV are shown in left, middle, and right columns, respectively.
	For the BB wave function, the isosceles triangle structure, $d_{1} = d_{2} \equiv d$, is assumed.}
	\label{fig:overlap_surface}
\end{figure*}

In Fig.~\ref{fig:overlap_surface}, we show overlap surfaces between the $0^{+}$ GCM states for $u_{\rm ls} = 0, 1600, 3200$ MeV and the single BB wave functions,
\begin{equation}
	O(d, \theta) = \langle \Psi_{0^{+}} | \Phi^{({\rm BB})} (d, \theta) \rangle.
\end{equation}
For the BB wave function, the isosceles triangle structure is assumed. 
These overlap surfaces show the cluster motion projected onto the 3$\alpha$ cluster configuration space for $0^{+}$ states.
The $u_{\rm ls} = 0$ MeV result for the weak limit of the spin-orbit force corresponds to the 3$\alpha$ cluster model calculation without $\alpha$-clusters breaking.
The $u_{\rm ls} = 3200$ MeV result for the strong limit of the spin-orbit force is the extreme case in which the $p_{3/2}$ subshell closure state comes down to the lowest state after level crossing over $3\alpha$ cluster states is completed, as described in the previous section.
The $u_{\rm ls} = 1600$ MeV result reproduces well the experimental results, as explained above, and is considered to be a reasonable result with the mixing of $\alpha$-cluster breaking. 
We discuss how $\alpha$-cluster breaking affects cluster structures in $0^{+}$ states by comparing the results with and without $\alpha$-cluster breaking, corresponding to $u_{\rm ls} = 0$ MeV and $u_{\rm ls} = 1600$ MeV, respectively.

First we compare the structure of the $0^{+}_{1}$ state with and without $\alpha$-cluster breaking.
For $u_{\rm ls} = 0$ MeV (without $\alpha$-cluster breaking), the wave function has a large overlap around the energy minimum in the $3\alpha$ cluster model space and also contains components of developed cluster configurations with larger inter-cluster distances than the energy minimum configuration.
For $ u_{\rm ls} = 1600$ MeV (with $\alpha$-cluster breaking), the behavior of the overlap surface is similar to that for $u_{\rm ls} = 0$ MeV.
However, the absolute amplitude around the energy minimum decreases and 3$\alpha$ cluster components become small because of the significant mixing of the $p_{3/2}$ subshell closure component in the $0^{+}_{1}$ state, as shown in Table.~\ref{table:squared_overlap_p3/2}.

Next we discuss $\alpha$-cluster breaking effects on the structure of the $0^{+}_{2}$ state, which reflects the structure change of the $0^{+}_{1}$ state because of the orthogonal condition to lower states.
For $u_{\rm ls} = 0$ MeV (without $\alpha$-cluster breaking), the overlap is distributed over a wide area of well-developed cluster structures, indicating that the $0^{+}_{2}$ state is described by the superposition of various triangle configurations of three $\alpha$ clusters. 
Consequently, it has a large rms radius, as explained above.
For $ u_{\rm ls} = 1600$ MeV (with $\alpha$-cluster breaking), the overlap is also distributed over a wide area of well-developed cluster structures and is qualitatively similar to the distribution for $u_{\rm ls} = 0$ MeV. 
However, quantitatively, the overlap shifts to smaller bending angles and the components for acute triangles are slightly enhanced. 
Moreover, the maximum peaks also shift toward shorter distances region compared to those for $u_{\rm ls} = 0$ MeV.

These results imply that the spatial development of the cluster structure is reduced in the $0^{+}_{2}$ state.
Namely, the gas-like feature of the $0^{+}_{2}$ state is slightly reduced by $\alpha$-cluster breaking, implying that $\alpha$-cluster breaking has the effect of attracting $\alpha$ clusters.
The differences in cluster structures in the $0^{+}_{2}$ states for $u_{\rm ls} = 1600$ MeV and for $u_{\rm ls} = 0$ MeV can be understood as follows.
Without $\alpha$-cluster breaking, the compact 3$\alpha$ cluster component in the $0^{+}_{2}$ is hindered due to orthogonality to the $0^{+}_{1}$ state, which has a compact $3\alpha$ cluster structure. 
With $\alpha$-cluster breaking, this hindrance is weakened because the 3$\alpha$ cluster component in the $0^{+}_{1}$ state decreases with the mixing of the non-cluster component of the $p_{3/2}$ subshell closure configuration. 

In the structure of the $0^{+}_{3}$ state, a qualitative difference is observed between the results with and without $\alpha$-cluster breaking.
For $u_{\rm ls} = 0$ MeV (without $\alpha$-cluster breaking), the phase of the overlap changes as $\theta$ increases along the line for $d \sim 5$ fm.
It has a negative minimum around  $(d, \theta) = (5.5 \text{ fm}, 60^{\circ})$ and a positive maximum $(d, \theta) = (4 \text{ fm}, 150^{\circ})$ with almost the same amplitude. 
This indicates that the $0^{+}_{3}$ state is the vibration mode of acute and obtuse triangle configurations.
For $u_{\rm ls} = 1600$ MeV (with $\alpha$-cluster breaking), the behavior differs from the vibration mode.
The amplitude of the negative minimum for the acute triangle configuration is considerably hindered, and as a result, the $0^{+}_{3}$ state is dominated by the obtuse triangle configuration.

The hindrance in the amplitude around $\theta = 60^{\circ}$ originates from orthogonality to the $0^{+}_{2}$ state.
As explained above, the components of acute triangle configurations increase in the $0^{+}_{2}$ state; therefore, the $0^{+}_{3}$ state loses components of acute triangle configurations due to orthogonality to the $0^{+}_{2}$ state. 
The components of obtuse triangle configurations in the $0^{+}_{2}$ state are relatively smaller than those of acute triangle configurations, and as a result, the overlap for the $0^{+}_{3}$ state concentrates on obtuse triangle configurations for a chain-like open triangle structure.
Because of $\alpha$-cluster breaking, the structure of the $0^{+}_{3}$ state changes from the vibration mode of acute and obtuse triangle configurations to the chain-like open triangle structure.
Hence, $\alpha$-cluster breaking produces significant effects not only in the ground state but also in the cluster configurations of excited states. 
This is surprising because the naive expectation is that $\alpha$-cluster breaking can affect only ground state properties.

\begin{figure}[tb]
	\centering
	\includegraphics[width = 8.6cm]{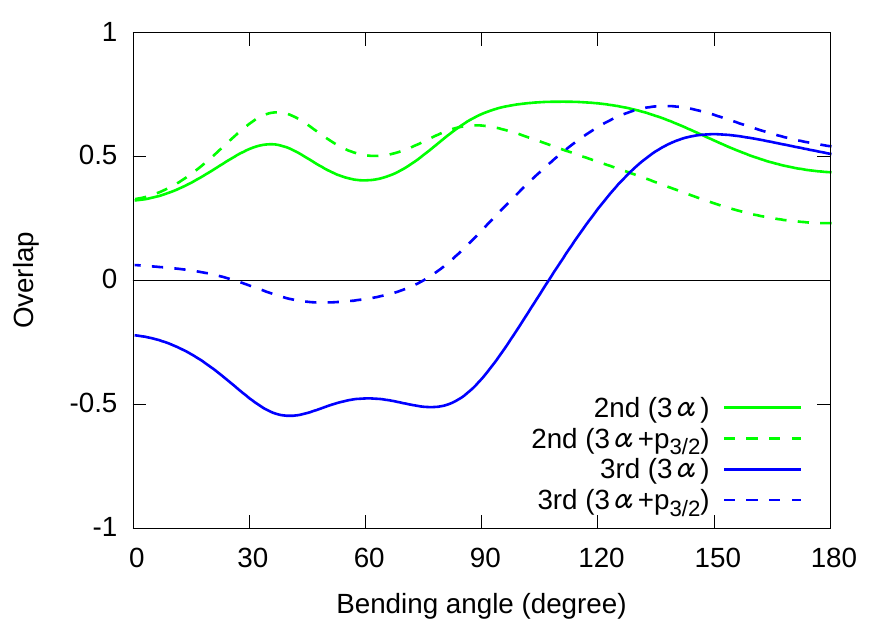}
	\caption{(color online).
	Overlap curves between $0^{+}$ GCM states and single BB wave functions for $u_{\rm ls} = 0$ MeV (3$\alpha$) and $u_{\rm ls} = 1600$ MeV (3$\alpha$+$p_{3/2}$) on the $d = 6.0 - 2.5 \theta / 180^{\circ} $ line.}
	\label{fig:overlap_2nd3rd}
\end{figure}

To compare in more detail the results with and without $\alpha$-cluster breaking for the $0^{+}_{2}$ and $0^{+}_{3}$ states, Fig.~\ref{fig:overlap_2nd3rd} shows a one-dimensional plot of the overlap of $0^{+}$ GCM states for $u_{\rm ls} = 0$ MeV (3$\alpha$) and $u_{\rm ls} = 1600$ MeV (3$\alpha$+$p_{3/2}$) with the single BB wave functions on the $d = 6.0 - 2.5 \theta / 180^{\circ} $ line.
In the 3$\alpha$ case, the overlap curve for the $0^{+}_{2}$ state is distributed almost uniformly over the entire region of the bending angle $\theta$.
To satisfy the orthogonal condition to the $0^{+}_{2}$ state, the overlap curve of the $0^{+}_{3}$ has a nodal structure with a node at $\theta = 100^{\circ}$ and almost the same heights of negative and positive amplitudes. 
In the 3$\alpha$+$p_{3/2}$ case, the overlap curve of the $0^{+}_{2}$ state leans toward the small $\theta$ region.
The overlap curve of the $0^{+}_{3}$ leans toward the large $\theta$ region, where chain-like open triangle configurations of three $\alpha$ clusters appear.

The $u_{\rm ls} = 3200$ MeV result for the strong limit of the spin-orbit force is the extreme case in which the $p_{3/2}$ subshell closure state comes down to the lowest state after the level crossing over $3\alpha$ cluster states is completed. 
Because the model space of 3$\alpha$ cluster configurations almost decouples from the $p_{3/2}$ subshell closure wave function in the ground state, 3$\alpha$ cluster structures appear in excited $0^{+}$ states. 
Consequently, the $0^{+}_{2}$, $0^{+}_{3}$, and $0^{+}_{4}$ have 3$\alpha$ cluster structures of the compact triangle, gas-like, and vibrational states; these are qualitatively similar to those of the $0^{+}_{1}$, $0^{+}_{2}$, and $0^{+}_{3}$ states in the $u_{\rm ls} = 0$ MeV results obtained without $\alpha$-cluster breaking, respectively. 
Strictly, the $p_{3/2}$ subshell closure wave function is not orthogonal to the 3$\alpha$ cluster model space, and its component in the ground state partially truncates the $3\alpha$ cluster model space; therefore, the consistency of cluster structures between $u_{\rm ls} = 0$ MeV and $u_{\rm ls} = 3200$ MeV is not entire.

We comment on the dependence on the width parameter of the $p_{3/2}$ subshell closure wave function.
In the present study, we adopted the same width parameter of the $p_{3/2}$ subshell closure wave function with the BB wave functions.
In the previous study of $^{12}$C with the method of variation after the spin-parity projection in the framework of AMD~\cite{en'yo_07}, very similar results to the present ones were obtained even though a different width parameter $\nu = 0.19$ fm$^{-2}$, which gives a larger radius for $p_{3/2}$ subshell closure wave function,  was adopted.
Therefore, we expect the results are not so sensitive to the choice of the width parameter, and even if we choose the different width parameter from the present study, we will obtain qualitatively similar results.
The further investigation using the different width parameters between the $p_{3/2}$ subshell closure wave function and BB wave functions is a future problem.

\subsection{Comparison with other models}

\begin{figure*}[tb]
	\centering
	\includegraphics[width = 0.9\linewidth]{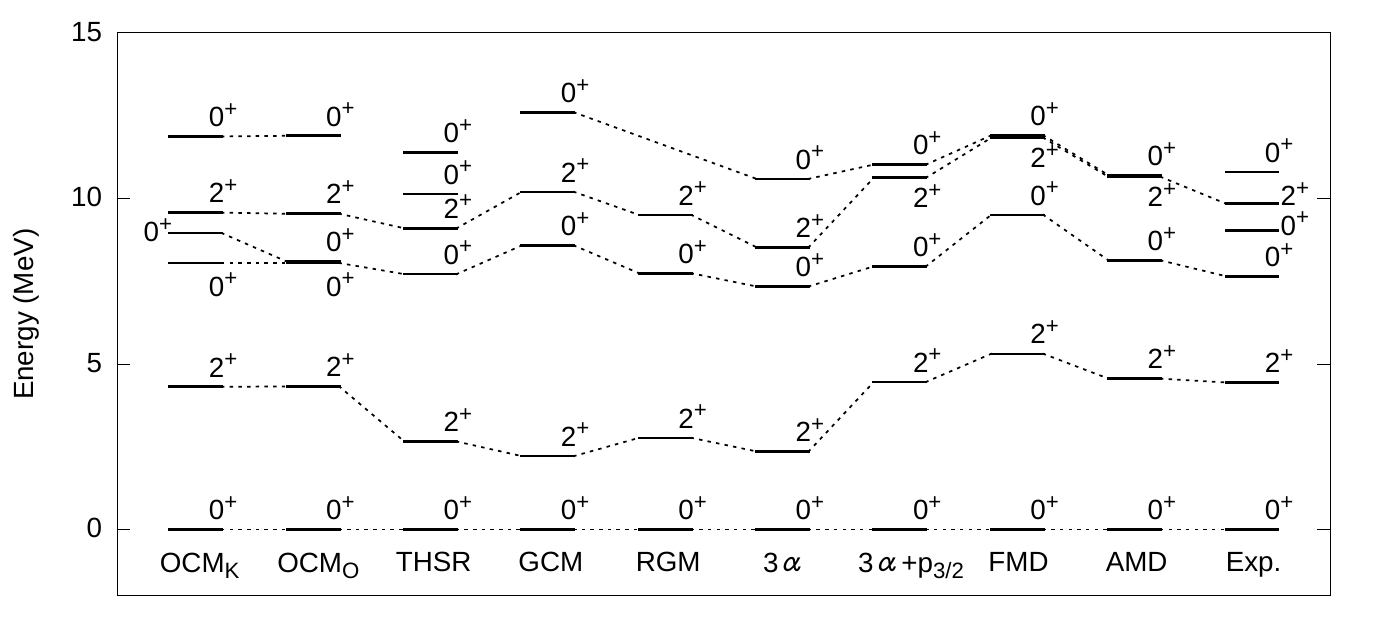}
	\caption{Comparison of levels from a semi-microscopic 3$\alpha$ cluster model (OCM$_{\text{K}}$~\cite{kurokawa_05, kurokawa_07} and OCM$_{\text{O}}$~\cite{ohtsubo_13}), microscopic 3$\alpha$ cluster models (THSR~\cite{funaki_14}, GCM~\cite{uegaki_77}, RGM~\cite{kamimura_77}, and 3$\alpha$), and models including $\alpha$-cluster breaking component (3$\alpha$+$p_{3/2}$, FMD~\cite{chernykh_07}, and AMD~\cite{en'yo_07}) with experimental data (Exp.~\cite{selove_90, zimmerman_13_2}).}
	\label{fig:comparison_levels}
\end{figure*}

In Fig.~\ref{fig:comparison_levels}, we compare energy levels obtained from the 3$\alpha$ and $3\alpha+p_{3/2}$ calculations of the present results with those from various theoretical models.
The first and second columns are the levels calculated by OCM~\cite{kurokawa_05, kurokawa_07,ohtsubo_13}, which is a semi-microscopic 3$\alpha$ cluster model.
The third, forth, and fifth columns are those calculated by the microscopic cluster models of THSR~\cite{funaki_14}, GCM~\cite{uegaki_77}, and resonating group method (RGM)~\cite{kamimura_77}, respectively.
In the sixth and seventh columns, the present results for $u_{\rm ls} = 0$ MeV (3$\alpha$) and $u_{\rm ls} = 1600$ MeV (3$\alpha$+$p_{3/2}$) for the microscopic calculations without and with $\alpha$-cluster breaking are shown, respectively.
The eighth and ninth columns are the levels calculated by FMD~\cite{chernykh_07} and AMD~\cite{en'yo_07}, which are microscopic twelve-body approaches without assuming the presence of any cluster and can contain $\alpha$-cluster breaking components.
The tenth column is experimental data~\cite{selove_90, zimmerman_13_2}.

The microscopic $\alpha$-cluster models underestimate the level spacing between the $0^{+}_{1}$ and $2^{+}_{1}$ states by about a factor of two, whereas models that include an $\alpha$-cluster breaking component reproduce it properly.
This is an evidence for $\alpha$-cluster breaking.
The OCM calculations fit this level spacing by hand adjustment of a phenomenological 3$\alpha$ potential.

Experimentally, four $0^{+}$ states have been reported. 
A newly measured $0^{+}$ state around 10 MeV is missing from theoretical calculations except for the OCM and THSR calculations.
Even though the number of $0^{+}$ states obtained in the OCM and THSR calculations is consistent with the experimental result, the calculated energy spectra for the $0^+_{2,3,4}$ and $2^{+}_{2}$ states around 10 MeV are not satisfactory. 
The relative energy position of the $2^{+}_{2}$ state to the $0^{+}$ states is very important for determining the band structure.
It is important to reproduce the experimental energy spectra for $0^{+}$ and $2^{+}$ states, including the new $0^{+}$ state, to clarify the structures of excited states near the threshold energy.

\begin{table*}[tb]
	\caption{Comparison of $E2$ transition strengths between $0^{+}$ and $2^{+}$ states.
	Labels are the same as in Fig.~\ref{fig:comparison_levels}.
	The unit is $e^{2}$fm$^{4}$.}
	\label{table:comparison_E2}
	\centering
	\begin{tabular}[t]{ccccccccc} \hline \hline
	& \multicolumn{6}{c}{Model} & \\ \cline{2-8}
	Transition 					& THSR 	& GCM 	& RGM 	& 3$\alpha$ 	& 3$\alpha$+$p_{3/2}$ 	& FMD 	& AMD 	& Exp. \\ \hline
	$2^{+}_{1} \rightarrow 0^{+}_{1}$ 	& 9.5 	& 8.0 	& 9.3 	& 10.8 		& 7.4 				& 8.69 	& 8.5 	& $7.6 \pm 0.4$ \\
	$2^{+}_{1} \rightarrow 0^{+}_{2}$ 	& 0.97 	& 0.7	& 1.1 	& 1.4 		& 5.1 				& 3.83 	& 5.1 	& $2.6 \pm 0.4$ \\
	$2^{+}_{1} \rightarrow 0^{+}_{3}$ 	& 		& 		& 		& 0.4 		& 0.2 				& 		& 		& \\
	$2^{+}_{2} \rightarrow 0^{+}_{1}$ 	& 2.4 	& 		& 2.5 	& 4.0 		& 1.1 				& 		& 0.4 	& 1.57$^{+0.14}_{-0.11}$ \\
	$2^{+}_{2} \rightarrow 0^{+}_{2}$ 	& 295 	& 		& 210 	& 183 		& 76.5 				& 		& 102 	& \\
	$2^{+}_{2} \rightarrow 0^{+}_{3}$ 	& 104 	& 		& 		& 64.4 		& 166 				& 		& 311 	& \\
	\hline \hline
	\end{tabular}
\end{table*}

In Table~\ref{table:comparison_E2}, we compare $E2$ transition strengths from the present results with those from several theoretical models.
The labels in the table are the same as those in Fig.~\ref{fig:comparison_levels}.
The microscopic $\alpha$-cluster models without $\alpha$-cluster breaking tend to overestimate the $2^{+}_{1} \rightarrow 0^{+}_{1}$ transition strengths, whereas models with $\alpha$-cluster breaking reproduce it properly. 
For the reproduction of the $2^{+}_{2} \rightarrow 0^{+}_{1}$ transition strength, the present 3$\alpha$+$p_{3/2}$ agrees with the experimental data better than the other models.

The $E2$ transition strengths from the $2^{+}_{2}$ state are important for determining the band assignment for this state, as discussed before. 
Although the $2^{+}_{2}$ state has strong $E2$ transitions to the $0^{+}_{2}$ and $0^{+}_{3}$ states, the major transition differs between calculations with and without $\alpha$-cluster breaking. 
The $2^{+}_{2} \rightarrow 0^{+}_{3}$ transition strength is significantly larger than the $2^{+}_{2} \rightarrow 0^{+}_{2}$ one in the $3\alpha+p_{3/2}$ and AMD calculations with $\alpha$-cluster breaking.
However, the relation of strengths is opposite and the $2^{+}_{2} \rightarrow 0^{+}_{2}$ transition is dominant in the 3$\alpha$ and THSR calculations without $\alpha$-cluster breaking. 
This implies that the band-head of the $2^{+}_{2}$ state is changed from the $0^{+}_{2}$ state to the $0^{+}_{3}$ state by $\alpha$-cluster breaking.
This difference between calculations with and without $\alpha$-cluster breaking comes from the difference of cluster configurations in the $0^{+}_{2}$ and $0^{+}_{3}$ states due to $\alpha$-cluster breaking.
As explained above, with $\alpha$-cluster breaking, the cluster development of the $0^{+}_{2}$ state is reduced to decrease the $E2$ transition from the $2^{+}_{2}$ state. 
Moreover, the structure of the $0^{+}_{3}$ state changes from the vibration mode of acute and obtuse triangle configurations of three $\alpha$ clusters to a chain-like open triangle structure.
Therefore, the phase with the $2^{+}_{2}$ becomes more coherent in the results with $\alpha$-cluster breaking.
We stress that $\alpha$-cluster breaking caused by the spin-orbit force has significant effects on transition strengths and changes band structure. 
This implies that, to investigate structures of excited cluster states in $^{12}$C, it is important to consider $\alpha$-cluster breaking, in particular, mixing of the $p_{3/2}$ subshell closure configuration in $0^{+}$ states.

\subsection{Linear chain-like band}

In the results with $\alpha$-cluster breaking, we assigned the $0^{+}_{3}$ and $2^{+}_{2}$ states as the band members because of the major $E2$ transition as discussed above.
Since the results with $\alpha$-cluster breaking reproduce the experimental data such as the large level spacing between $0^{+}_{1}$ and $2^{+}_{1}$ states, rms-radii, and $E2$ transition strengths between low-lying states well, we consider the $0^{+}_{3}$ state is likely to be the band-head of the $2^{+}_{2}$, although there is no experimental data for the $E2$ transitions from the $2^{+}_{2}$ state to the $0^{+}_{2}$ state and the $0^{+}_{3}$ state.

Strictly speaking, the structure of the $2^{+}_{2}$ state is different from that of the $0^{+}_{3}$ state because of 16\% of the $\alpha$-cluster breaking component in the $0^{+}_{3}$ state (See Table~\ref{table:squared_overlap_p3/2}).
It means that they are not an ideal rotational band constructed from an intrinsic state.
However, when we project the structures on the 3$\alpha$ cluster configuration space, we found that both of $0^{+}_{3}$ and $2^{+}_{2}$ states have maximum amplitude around obtuse triangle region.
Considering these points, the stronger $E2$ transition strength and maximum amplitude around the chain-like open triangle structure, this band can be considered a linear chain-like band.

In this band, the energy position of the $0^{+}_{3}$ state is slightly higher than that of the $2^{+}_{2}$ state in the present calculation with $\alpha$-cluster breaking.
The order of the $0^{+}$ state and the $2^{+}$ state is reverse to an ordinary rotational band. 
This may come from the mixing of the intrinsic structure of the $0^{+}_{2}$ state in the $2^{+}_{2}$ state, though the dominant component of the $2^{+}_{2}$ state is the intrinsic structure of the $0^{+}_{3}$ state. 
This can be understood from that the $E2$ transition from the $2^{+}_{2}$ state is not weak to the $0^{+}_{2}$ state though the strength is relatively smaller than that to the $0^{+}_{3}$ state.

\section{Summary}\label{summary}

To clarify $\alpha$-cluster breaking effects on 3$\alpha$ cluster structures in $^{12}$C, we investigated $^{12}$C with a hybrid model composed of the BB cluster model and the $p_{3/2}$ subshell closure wave function.
We superposed 252 BB wave functions to sufficiently describe the relative motions of clusters. 
The $\alpha$-cluster breaking caused by the spin-orbit force affects $0^{+}$ states through the mixing of the $p_{3/2}$ subshell closure wave function.
The present result reproduces well the experimental results with the significant mixing of $\alpha$-cluster breaking. 
We have found that $\alpha$-cluster breaking significantly changes the cluster structures of $0^{+}$ states through orthogonality to lower states.
As a result of the structure changes of $0^{+}$ states, the band assignment for the $2^{+}_{2}$ state is changed.
To investigate $\alpha$-cluster breaking effects, we compared 3$\alpha$ configurations contained in $0^{+}$ states obtained from calculations with and without $\alpha$-cluster breaking.

In calculations with and without $\alpha$-cluster breaking, the $0^{+}_{1}$ state has a compact $3\alpha$ cluster structure as the dominant component and also contains components of developed cluster configurations.
However, in the calculation with $\alpha$-cluster breaking, 3$\alpha$ cluster components becomes relatively small because of significant mixing of the $p_{3/2}$ subshell closure component.

In both the calculations, the $0^{+}_{2}$ state is described by the superposition of various triangle configurations of three $\alpha$ clusters.
However, in the calculation with $\alpha$-cluster breaking, the spatial development of the cluster structure and the gas-like feature of the $0^{+}_{2}$ state are slightly reduced.
Orthogonality to the $0^{+}_{1}$ state cannot hinder the $0^{+}_{2}$ state including more compact $\alpha$ cluster components because the 3$\alpha$ cluster component in the $0^{+}_{1}$ state decreases with $\alpha$-cluster breaking. 

Because of $\alpha$-cluster breaking, the structure of the $0^{+}_{3}$ state changes from the vibration mode of acute and obtuse triangle configurations to the chain-like open triangle structure.
The components of acute triangle configurations increases in the $0^{+}_{2}$ state; therefore, the $0^{+}_{3}$ state loses components of acute triangle configurations because of orthogonality to the $0^{+}_{2}$ state. 

Because of structure changes of $0^{+}$ states by $\alpha$-cluster breaking, transition strengths change significantly.
The band assignment for the $2^{+}_{2}$ state differs between calculations with and without $\alpha$-cluster breaking.
Namely, in the model calculation without $\alpha$-cluster breaking, the $0^{+}_{2}$ state is assigned to be the band-head of the $2^{+}_{2}$ state.
However, when we incorporate $\alpha$-cluster breaking caused by the spin-orbit force, the $0^{+}_{3}$ state is regarded as the band-head of the $2^{+}_{2}$ state.
The $0^{+}_{3}$ state is likely to be the band-head of the $2^{+}_{2}$ state instead of the $0^{+}_{2}$ state because the present calculation reproduces experimental data of the low-lying states well with $\alpha$-cluster breaking.
Considering the stronger $E2$ transition strength and maximum amplitude around the chain-like open triangle structure, this band can be considered a linear chain-like band.
Unfortunately, there is no experimental data for the $E2$ transitions from the $2^{+}_{2}$ state to excited $0^{+}$ states. 
Moreover, experimental information about two $0^+$ states around 10 MeV is not enough to assign the theoretical $0^{+}_{3}$ state to either of the experimental $0^{+}$ states.
Further experimental data related to $0^{+}$ and $2^{+}$ states in this energy region are required to clarify the band assignment of these states.

We stress that $\alpha$-cluster breaking caused by the spin-orbit force gives significant effect on cluster structures, transition strengths, and band structure. 
To investigate structures of excited cluster states in $^{12}$C, it is important to consider $\alpha$-cluster breaking, in particular, mixing of the $p_{3/2}$ subshell closure configuration in $0^{+}$ states.

\section*{Acknowledgments} 
The computational calculations of this work were performed by the supercomputers at YITP.
This work was supported by JSPS KAKENHI Grant Nos. 25887049 and 26400270.


\begin{thebibliography}{99}
\bibitem{ikeda_68}
	K. Ikeda, N. Takigawa, and H. Horiuchi, Prog. Theor. Phys. Suppl. {\bf E68}, 464 (1968).
\bibitem{itagaki_04}
	N. Itagaki, S. Aoyama, S. Okabe, and K. Ikeda, Phys. Rev. C {\bf 70}, 054307 (2004).
\bibitem{en'yo_98}
	Y. Kanada-En'yo, Phys. Rev. Lett. {\bf 81}, 5291 (1998).
\bibitem{neff_04}
	T. Neff and H. Feldmeier, Nucl. Phys. {\bf A738}, 357 (2004).
\bibitem{en'yo_07}
	Y. Kanada-En'yo, Prog. Theor. Phys. {\bf 117}, 655 (2007).
\bibitem{chernykh_07}
	M. Chernykh, H. Feldmeier, T. Neff, P. von Neumann-Cosel, and A. Richter, Phys. Rev. Lett. {\bf 98}, 032501 (2007).
\bibitem{en'yo_12}
	Y. Kanada-En'yo, M. Kimura, and A. Ono, Prog. Theor. Exp. Phys. {\bf 2012}, 01A202 (2012).
\bibitem{fukuoka_13}
	Y. Fukuoka, S. Shinohara, Y. Funaki, T. Nakatsukasa, and K. Yabana, Phys. Rev. C {\bf 88}, 014321 (2013).
\bibitem{uegaki_77}
	E. Uegaki, S. Okabe, Y. Abe, and H. Tanaka, Prog. Theor. Phys. {\bf 57}, 1262 (1977).
\bibitem{kamimura_77}
	Y. Fukushima and M. Kamimura, \textit{in Proceedings of the International Conference on Nuclear Structure, Tokyo, 1977}, edited by T. Marumori [J. Phys. Soc. Jpn. {\bf 44}, 225 (1978)].
\bibitem{tohsaki_01}
	A. Tohsaki, H. Horiuchi, P. Schuck, and G. R\"{o}pke, Phys. Rev. Lett. \textbf{87}, 192501 (2001).
\bibitem{funaki_03}
	Y. Funaki, A. Tohsaki, H. Horiuchi, P. Schuck, and G. R\"{o}pke, Phys. Rev. C \textbf{67}, 051306(R) (2003).
\bibitem{funaki_05}
	Y. Funaki, A. Tohsaki, H. Horiuchi, P. Schuck, and G. R\"{o}pke, Eur. Phys. J. A \textbf{24}, 321 (2005).
\bibitem{suhara_14}
	T. Suhara, Y. Funaki, B. Zhou, H. Horiuchi, and A. Tohsaki, Phys. Rev. Lett. \textbf{112}, 062501 (2014).
\bibitem{funaki_14}
	Y. Funaki, arXiv:1408.5855 [nucl-th] and private communication.
\bibitem{yamada_08}
	T. Yamada, Y. Funaki, H. Horiuchi, K. Ikeda, and A. Tohsaki, Prog. Theor. Phys. {\bf 120}, 1139 (2008).
\bibitem{navratil_03}
	P. Navr\'{a}til and W. E. Ormand, Phys. Rev. C {\bf 68}, 034305 (2003).
\bibitem{navratil_09}
	P. Navr\'{a}til, S. Quaglioni, I. Stetcu, and B. R. Barrett, J. Phys. G: Nucl. Part. Phys. {\bf 36} 083101 (2009).
\bibitem{neff_08}
	T. Neff and H. Feldmeier, J. Phys. Conf. Ser.. {\bf 111} 012007 (2008).
\bibitem{epelbaum_12}
	E. Epelbaum, H. Krebs, T. A. L\"{a}hde, D. Lee, U. Mei{\ss }ner, Phys. Rev. Lett. {\bf 109}, 252501 (2012).
\bibitem{suhara_10}
	T. Suhara and Y. Kanada-En'yo, Prog. Theor. Phys. {\bf 123}, 303 (2010).
\bibitem{freer_09}
	M. Freer {\it et al}, Phys. Rev. C {\bf 80}, 041303(R) (2009).
\bibitem{itoh_11}
	M. Itoh {\it et al},. Phys. Rev. C {\bf 84}, 054308 (2011).
\bibitem{zimmerman_13_1}
	W. R. Zimmerman {\it et al}, Phys. Rev. Lett. {\bf 110}, 152502 (2013). 
\bibitem{zimmerman_13_2}
	W. R. Zimmerman, ``Direct Observation of the Second 2$^{+}$ State in $^{12}$C'' (2013). Doctoral Dissertations. Paper 230.
\bibitem{itoh_13}
	M. Itoh {\it et al}, J. Phys. Conf. Ser. {\bf 436}, 012006 (2013).
\bibitem{kurokawa_05}
	C. Kurokawa and K. Kat\={o}, Phys. Rev. C {\bf 71}, 021301 (2005).
\bibitem{kurokawa_07}
	C. Kurokawa and K. Kat\={o}, Nucl. Phys. {\bf A792}, 87 (2007).
\bibitem{ohtsubo_13}
	S. Ohtsubo, Y. Fukushima, M. Kamimura, and E. Hiyama, Prog. Theor. Exp. Phys. {\bf 2013}, 073D02 (2013).
\bibitem{brink_66}
	D. M. Brink, {\it Proc. Int. School of Physics Enrico Ferm, Course 36}, Varenna, ed. C. Bloch (Academic Press, New York, 1966).
\bibitem{okabe_79}
	S. Okabe and Y. Abe, Prog. Theor. Phys. {\bf 61}, 1049  (1979).
\bibitem{itagaki_05}
	N. Itagaki, H. Masui, M. Ito, and S. Aoyama, Phys. Rev. C {\bf 71} 064307 (2005). 
\bibitem{suhara_13}
	T. Suhara, N. Itagaki, J. Cseh, and M. P{\l}oszajczak, Phys. Rev. C {\bf 87}, 054334 (2013).
\bibitem{volkov_65}
	A. Volkov, Nucl. Phys. {\bf 74}, 33 (1965).
\bibitem{yamaguchi_79}
	N. Yamaguchi, T. Kasahara, S. Nagata, and Y. Akaishi, Prog. Theor. Phys. {\bf 62}, 1018 (1979).
\bibitem{selove_90}
	F. Ajzenberg-Selove and J. H. Kelley, Nucl. Phys. {\bf A506}, 1 (1990).
\bibitem{ozawa_01}
	A. Ozawa, T. Suzuki, and I. Tanihata, Nucl. Phys. {\bf A693}, 32 (2001).
\bibitem{danilov_09}
	A.N. Danilov, T.L. Belyaeva, A.S. Demyanova, S.A. Goncharov, and A.A. Ogloblin, Phys. Rev. C {\bf 80}, 054603 (2009).

\end{thebibliography}
\end{document}